\documentclass[aps,prb,twocolumn]{revtex4}

\usepackage[dvips]{graphicx}
\usepackage{amssymb,amsfonts,amsmath}
\usepackage{eucal,bm}
\usepackage{gensymb,subfigure}
\usepackage{natbib}
\usepackage{booktabs}
\usepackage{times,color,url}

\begin{document}

\title{Fault Induced Delayed Voltage Recovery\\ in a Long Inhomogeneous Power Distribution Feeder}

\author{Irina Stolbova $^{a}$}
\author{Scott Backhaus $^{b,c}$}
\author{Michael Chertkov $^{d,c}$}

\affiliation{$^a$ Moscow Institute of Physics and Technology, Dolgoprudnyj, Moscow Regiona 141700, Russia}
\affiliation{$^b$ Materials, Physics \& Applications Division, Los Alamos National Laboratory, NM 87545, USA}
\affiliation{$^c$ New Mexico Consortium, Los Alamos, NM 87544, USA}
\affiliation{$^d$ Theoretical Division and Center for Nonlinear Studies, Los Alamos National Laboratory, NM 87545, USA}

\date{\today}

\begin{abstract}
We analyze the dynamics of a distribution circuit loaded with many induction motor and subjected to sudden changes in voltage at the beginning of the circuit.  As opposed to earlier work \cite{13DCB}, the motors are disordered, i.e. the mechanical torque applied to the motors varies in a random manner along the circuit.  In spite of the disorder, many of the qualitative features of a homogenous circuit persist, e.g. long-range motor-motor interactions mediated by circuit voltage and electrical power flows result in coexistence of the spatially-extended and propagating normal and stalled phases.  We also observed a new phenomenon absent in the case without inhomogeneity/disorder.  Specifically, transition front between the normal and stalled phases becomes somewhat random, even when the front is moving very slowly or is even stationary.  Motors within the blurred domain appears in a normal or stalled state depending on the local configuration of the disorder. We quantify effects of the disorder and discuss statistics of distribution dynamics, e.g. the front position and width, total active/reactive consumption of the feeder and maximum clearing time.
\end{abstract}

\pacs{}
\keywords{Power Systems Dynamics|Voltage Collapse|Phase Transitions|Hysteresis}

\maketitle

\section{Introduction}

The majority of transient and dynamical stability studies in power systems focuses on high voltage transmission grids where detailed models of the generators and transmission lines are used. In contrast, these studies use crude aggregations of individual small loads to model aggregate load dynamics. Recent years have seen an increased emphasis on dynamical load models \cite{10Les,09FIDVR} for several reasons. First, transmission networks are being pushed harder and operated closer to their dynamical stability limits, and the uncertainty introduced by inaccurate linear dynamical load models presents an unacceptable operating risk \cite{02PKMDUZ}.  Second, collective nonlinear dynamical load behaviors such as induction motor stalling and Fault-Induced Delayed Voltage Recovery (FIDVR) are being excited by seemingly typical transmission grid behavior such as normal fault clearing \cite{92WSD,97Sha}.

In FIDVR, a short-lived but significant perturbation created by a transmission fault synchronizes the dynamical behavior of a large fraction of individual induction motor loads, and the ensuing collective dynamics induce a voltage collapse-like event that propagates to the transmission grid.  Complicating the situation further, the future will likely see the introduction of many new ``smart'' consumer devices that include local controllers making decisions based upon local measurements.  The actions of these independent and potentially stochastic load controls will yield new dynamics.  There are no tools that can predict the collective effect of these dynamics and the potential impact on transmission grids. The example of FIDVR demonstrates the importance of understanding and modeling emergent collective effects in distribution dynamics for modeling of transmission grid dynamics. This work extends a model of distribution dynamics \cite{13DCB,11CBTCL} to study the impact of inhomogeneity in distribution circuit loading on the FIDVR dynamics mentioned above.

Diversity of distribution circuits makes detailed, component-by-component modeling of the many configurations very difficult.  Even if modeling of the family of such configurations were performed, it is not clear that the results could be understood or displayed in way that enables intuitive interpretation and understanding. Instead, we adopt the model of \cite{11CBTCL} and \cite{13DCB} where the distribution circuit power flows and load dynamics are represented as a continuum system and model the spatiotemporal dynamics using Partial Differential Equations (PDE).  Similar approaches have been used to model dynamical effects in transmission grids \cite{Thorp1998,Seyler2004}.  This PDE approach reveals the nontrivial interplay of the dynamics of loads via the spatial coupling provided by power flows over the electrical network.  In \cite{13DCB}, this approach was used to reveal the qualitative behavior of FIDVR dynamics in a uniformly-loaded distribution circuit, and in \cite{12WTC} this approach was used to explore new equilibrium states of a distribution circuit with a spatially-uniform installation of actively-controlled photovoltaic inverters.  In both of these cases, the structure of the PDE reveals the important long-range interactions between local load behavior (dynamic or static) mediated by the power flows along the distribution circuit.

In this manuscript, we extend the work of \cite{13DCB} by investigating the effects of spatially inhomogeneous induction motor loading.  Variation in loading is expected to modify the behavior of FIDVR dynamics, however, both the qualitative and quantitative effects are not clear.  Pockets of high or low induction motor load could locally arrest or enhance the propagation of a FIDVR event, but long-range effects are also possible.  We create realizations of circuit loadings using spatially-correlated Gaussian distribution of load parameters and study the behavior of the FIDVR dynamics as a function of the amplitude of the load disorder and the correlation length.  From these studies, we find that, at least for relatively low amplitude and spatially short-correlated disorder, the effects on FIDVR dynamics are local.  This conclusion has important implications for control, specifically, that a control scheme for eliminating or correcting disruptive FIDVR events is not strongly dependent on the details of the induction motor loading.

The rest of the manuscript is organized as follows.  Section~\ref{sec:Tech intro} introduces the PDE model of distribution dynamics and load inhomogeneity and briefly describes the numerical methods used to integrate the PDE model. Section~\ref{sec:Results} presents the results of numerically integrating the PDE model with different types of load inhomogeneity, parameterized by the amplitude and the correlation length of the disorder, on different dynamical transients. Finally, Section~\ref{sec:Conclusions} provides some conclusions and potential areas for future work.

\section{Technical Introduction}\label{sec:Tech intro}

\subsection{Dynamics of an Individual Motor}

\begin{figure}
	\centering
	\includegraphics[width=.9\linewidth]{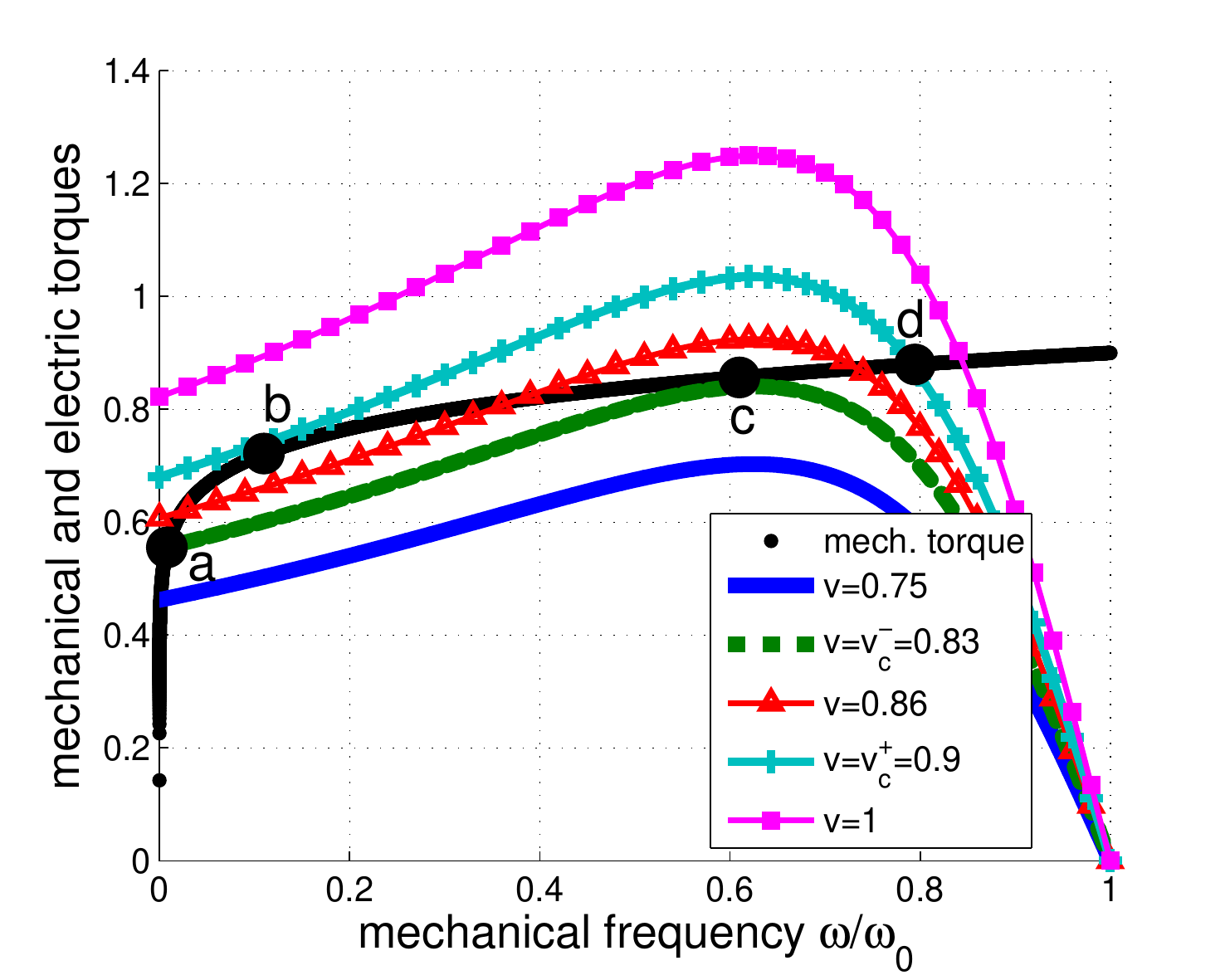}
	\caption{
Electric and mechanical torques as functions of the mechanical frequency $\omega/\omega_0$ for a range of motor terminal voltages $v$, reference mechanical torque $T_0=0.32$, and $\alpha=0.1$.  For the $v=0.86$ electrical torque curve (red triangles), there are three equilibrium solutions indicated by intersections with the mechanical torque curve (black solid line).  The solution with the highest $\omega/\omega_0$ is the ``normal'' stable solution with the induction motor rotating near the grid frequency $\omega_0$.  The ``stalled-state'' with $\omega/\omega_0\simeq 0$ is also stable while the intermediate solution is unstable.  For $v>v_c^+=0.9$ (light blue line with plus signs), there is only one solution corresponding to the normal state.  For $v<v_c^-=0.83$ (green dashed line), there is only one solution corresponding to the stalled state.  The points ($a,b,c,d$) correspond to the same labels in Fig.~\ref{fig:hysteresis-single}.}
	\label{fig:torques}
\end{figure}

We adopt the model of induction motor load and dynamics from \cite{98PHH}.  Here, we only describe the features of inductions motors that are important for the remainder of this manuscript. The real ($P$) and reactive ($Q$) powers drawn by and induction motor are
\begin{eqnarray}
&& P=\frac{s R_m v^2}{R^2_m+s^2 X^2_m},
\label{p-motor}\\
&& Q=\frac{s^2 X_m v^2}{R^2_m+s^2 X^2_m},
\label{q-motor-stat}
\end{eqnarray}
where $s=1-\omega/\omega_0$ is the slip of the motor's rotational frequency $\omega$ relative to the grid frequency $\omega_0$ ($0\leq s<1$); $v$ is the voltage at the motor terminals; and $X_m$ and $R_m$ are internal reactance and resistance of the motor, respectively.  Typically, $R_m/X_m =0.1\div 0.5$.

\begin{figure}
	\centering
	\includegraphics[width=.9\linewidth]{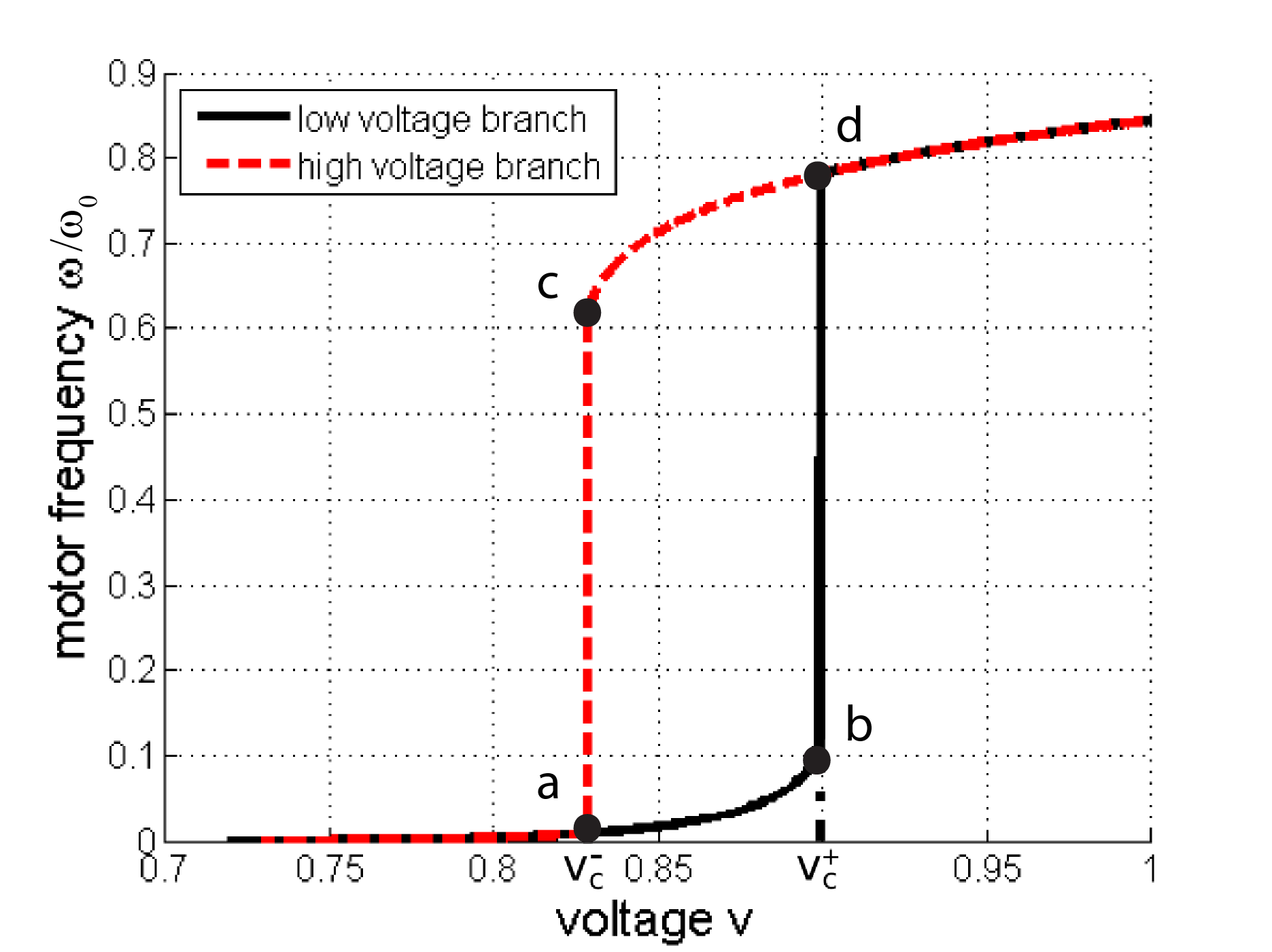}
	\caption{Hysteretic behavior of an induction motor stable steady states in Fig.~\ref{fig:torques} as the voltage $v$ at its terminals is varied.  The dashed red (solid black) curves indicate the path of equilibrium states as the voltage $v$ is decreased (increased) starting from the high-voltage normal (low-voltage stalled) state.  The vertical lines at the spinodal voltages $v_c^\pm$ indicated the abrupt hysteretic transitions between states.  $v_c^\pm$ correspond to the same labels in the legend of Fig.~\ref{fig:torques}. The points ($a,b,c,d$) correspond to the states where the motor must make transitions from normal to stalled ($c\rightarrow a$) and from stalled to normal $(b \rightarrow d)$.}
	\label{fig:hysteresis-single}
\end{figure}

The real power load $P$ creates an electric torque on the induction motor shaft which is countered by a mechanical load torque.  Any imbalance between these torques results in an angular acceleration of the motor's rotational inertia $M$, i.e.
\begin{eqnarray}
M\frac{d}{dt}\omega=\frac{P}{\omega_0}-T_0\left(\frac{\omega}{\omega_0}\right)^\alpha.
\label{omega-relaxation}
\end{eqnarray}
Here $T_0$ is a reference mechanical torque and $\alpha$ is indicative of different types of mechanical loads with $\alpha=1$ typical of fan loads and $\alpha<1$ typical of air-conditioning or other compressor loads. If $\alpha<\alpha_c \simeq 1$ and $T_0$ are fixed, there exist three steady solutions of Eqs.~(\ref{p-motor},\ref{omega-relaxation}) when $v$ is in a range between two spinodal voltages $v_c^-$ and $v_c^+$.  Figure~\ref{fig:torques} displays the mechanical torque (black curve) and the electrical torque (colored curves) for different values of $v$.  For a mid-range voltage of 0.86 (red triangles), the mechanical and electrical torque curves intersect for three values of $\omega/\omega_0$ defining three steady solutions.  For the high and low $\omega/\omega_0$ solutions, the torque balance for small deviations away from steady solution push $\omega/\omega_0$ back to the steady state.  The opposite is true for the mid-range $\omega/\omega_0$ solution making it unstable to small deviations.

From Fig.~\ref{fig:torques}, it is clear that small changes in $v$ in the vicinity of $v_c^+$ and $v_c^-$ can lead to drastic and hysteretic changes in $\omega$ resulting in a large changes in the motor's $P$ and $Q$ (via Eqs.~\ref{p-motor} and \ref{q-motor-stat}).  These hysteretic changes will be coupled back to the dynamics in Eq.~(\ref{omega-relaxation}) via the power flow equations in Section~\ref{sec:cont_model} \cite{13DCB}.  This hysteresis and coupling can be affected by inhomogeneity in loading, and these effects are explored int he remainder of this manuscript.

The effect of the hysteresis is cleanly displayed in Fig.~\ref{fig:hysteresis-single} where the $\omega/\omega_0$ for two stable steady states is plotted versus the motor terminal voltage $v$. Starting  in the high-voltage normal state (say $v\sim 1$), $v$ can be slowly decreased along the dashed red curve passing through state $d$. Further decreasing $v$ to state $c$, the ``normal'' state (i.e. the state with high $\omega/\omega_0$ suddenly disappears and the motor makes a transition to the ``stalled'' state at $a$.  Similarly, if we start from the low-voltage stalled state (say $v\sim 0.75$ on the black curve) and $v$ is increased slowly through state $a$ to $b$, the stalled state disappears and the motor makes a transition to the normal state at $d$.  For reference, the states ($a,b,c,d$) are also marked in Fig.~\ref{fig:torques}, and the same hysteresis loop can be traced out there.  The voltages $v_c^+$ and $v_c^-$ depend on induction motor parameters, and disorder in these parameters will result in neighboring segments of the circuit making state transitions at different times.  The discreteness of the transitions and the large changes in $\omega/\omega_0$, $P$, and $Q$ will significantly amplify the even a small amount of disorder.

\subsection{Continuum Model of Distribution Dynamics}\label{sec:cont_model}

The derivation of the continuous form of the DistFlow equations is described in \cite{11CBTCL}.  Here, we only summarize the results that are important to the rest of this manuscript.  The evolution of the real ($\rho$) and reactive ($\phi$) line flows is caused by loads or line losses, i.e.
\begin{eqnarray}
&& \partial_z \rho=-p-r\frac{\rho^2+\phi^2}{v^2},\label{p_cont}\\
&& \partial_z \phi=-q-x\frac{\rho^2+\phi^2}{v^2}. \label{q_cont}
\end{eqnarray}
Here $z$ is the coordinate along the distribution circuit, $r$, $x$ are the per-unit-length resistance and reactance densities of the lines (assumed independent of $z$) and $p(z)$ and $q(z)$ are the local densities of real and reactive powers consumed by the density of the spatially continuous distribution of motors \cite{11CBTCL} at the position $z\in[0;L]$.  The power flows $\rho$ and $\phi$ are related to the voltage at the same position according to \cite{11CBTCL}
\begin{eqnarray}
\partial_z v=-\frac{r \rho+ x \phi}{v}. \label{v_eq}
\end{eqnarray}

The load densities $p(z)$ and $q(z)$ in Eqs.~(\ref{p_cont},\ref{q_cont}) are related to $\omega(z)$ and $v(z)$ through the density versions of Eqs.~(\ref{p-motor},\ref{q-motor-stat},\ref{omega-relaxation})
\begin{eqnarray}
\mu\frac{d}{dt}\omega&=&\frac{p}{\omega_0}-\tau_0\left(\frac{\omega}{\omega_0}\right)^\alpha,
\label{torque-cont}\\
p&=&\frac{s r_m v^2}{r^2_m+s^2 x^2_m},
\label{p-motor-cont}\\
q&=&\frac{s^2 x_m}{r^2_m+s^2 x^2_m}v^2, 
\label{q-motor-cont}
\end{eqnarray}
where the conversion to continuous form consists of replacing $X_m,R_m$ and $P,Q,T_0,M$ by the respective densities $x_m$, $r_m$ and $p$, $q$, $\tau_0$, and $\mu$.  The new boundary conditions are
\begin{eqnarray}
v(0)=v_0,\quad \rho(L)=\phi(L)=0.
\label{bc}
\end{eqnarray}
Eqs.~(\ref{p_cont}-\ref{bc}) form our PDE model of a distribution feeder loaded with induction motors.

\subsection{Model of Disorder}

Analysis in \cite{13DCB} assumed that all of the induction motor parameters $\tau_0$ are constant, i.e. the circuit is uniformly loaded with identical induction motors all serving identical loads.  However, loading in distribution circuits is inhomogeneous and variable depending on, e.g., the time of day or environmental conditions.  We relax the assumption of uniform loading by introducing load inhomogeneity by making $\tau_0$ a random Gaussian variable centered on $\bar{\tau}_0$ (i.e. $\bar{\tau}_0=\mathbb{E}\left[\tau_0(z)\right]$).  Deviations from the mean $\delta(z)=\tau_0(z)-\bar{\tau}_0$ are statistically homogeneous with the covariance, $\mathbb{E}\left[\delta(z_1)\delta(z_2)\right]=(\tau_0\Delta)^2\exp(-|z_1-z_2|/z_d)$. The amplitude of the the disorder, as well as the correlation scale of the disorder are assumed small, $\Delta\ll 1, z_d/L\ll 1$, where $L$ stands for the length of the feeder. The Gaussian model is the simplest and most natural spatially smooth and two parametric (amplitude and correlation length) model of the disorder. To implement the Gaussian finite correlated model of the inhomogeneity/disorder in the simulations of Eqs.~(\ref{p_cont}-\ref{bc}), one sets up the spatial step size which is much smaller than the disorder's correlation length, $z_d$. The spatial step used in the simulation was $2.5*10^{-4}$ vs. $5*10^{-3}$ used for the minimum value of the correlation length.

\subsection{Numerical Simulation Approach}\label{sec:numerical_approach}

Eqs.~(\ref{p_cont}-\ref{bc}) are integrated using the following iterative procedure (see \cite{13DCB} for details). For given current values of $p$ and $q$, the discretized version of Eqs.~(\ref{p_cont},\ref{q_cont},\ref{v_eq}) are solved by a shooting method, i.e. using the fixed $v_0$ from Eq.~\ref{bc}, $\rho(0)$ and $\phi(0)$ are adjusted until the spatial integration of the time-independent equations accurately recreates the boundary conditions in Eq.~\ref{bc} at $z=L$.  Using this voltage profile, the motors' $\omega$ are updated using a time-discretized version of Eq.~(\ref{torque-cont}). Using the new values of $\omega$, the parameters $s$, $p$ and $q$ are updated using Eqs.~(\ref{p-motor-cont},\ref{q-motor-cont}) and the process repeats.
In the simulations we assume that $\alpha=0.1$, $x_m=0.4$, $r_m=0.15$, $r=0.5$, $x=0.5$, $L=0.5$, $\mu=0.1$, $w_0=1$ and $\bar{\tau}_0=0.9$.

\section{Numerical Experiments: Results}\label{sec:Results}

Eqs.~(\ref{p_cont}-\ref{bc}) are studied numerically under a range conditions:
\begin{itemize}
\item[(S)] Steady-state conditions---a constant $v_0$=1 is applied.
\item[(A)] Stalling front dynamics---starting from the previous steady-state conditions, $v_0$ is suddenly lowered by a range of $\Delta v$'s and the dynamics of the motor stalling front is studied.
\item[(B)] Restoration front dynamics---starting from a fully stalled condition (i.e. $v_0<1$ and all motors on the low voltage branch of Fig.~\ref{fig:hysteresis-single}), $v_0$ is suddenly raised to $1.0$, and the dynamics of the motor restoration front is studied.
\item[(C)] Fault-clearing dynamics---A fault clearing perturbation is emulated by combining case A and case B.  Starting from the steady-state condition from S, $v_0$ is lowered by $\Delta v$ for time $\tau_{cl}$ to create a stalling front.  After $\tau_{cl}$, $v_0$ is subsequently restored to 1.0, and the restoration is studied.  Our goal is to determine how the disorder in $\tau_0$ affects the maximum fault clearing time, i.e. the longest the fault can stay on while the all of the motors on the circuit recover to the high-voltage branch in Fig.~\ref{fig:hysteresis-single}.
\end{itemize}

In each study, statistics are gathered over many samples of the distribution of motor and feeder disorder, i.e. $\Delta$ and $z_d$.  For each simulated sample in case A and B, the following data is analyzed:
\begin{itemize}
\item The active and  reactive power flows at $z=0$, $\rho(0)$ and $\phi(0)$;
\item The position of the front $z_f$ and the width the front, $z_w$.
\end{itemize}
For case (C), the maximum fault clearing time $\tau_{cl}$ is found that still allows the circuit to recover to a fully normal state.  The goal of these studies is to quantify the effect of disorder effect on statistics of these data.  Aimed to test sensitivity of the results to the parameters of the disorder, $\Delta$ and $z_d$, were varied in our numerical experiments.

\subsection{Case S---``Normal'' Steady-State Solution}\label{sec:Case S}

\begin{figure}
	\centering	
\includegraphics[width=0.95\linewidth]{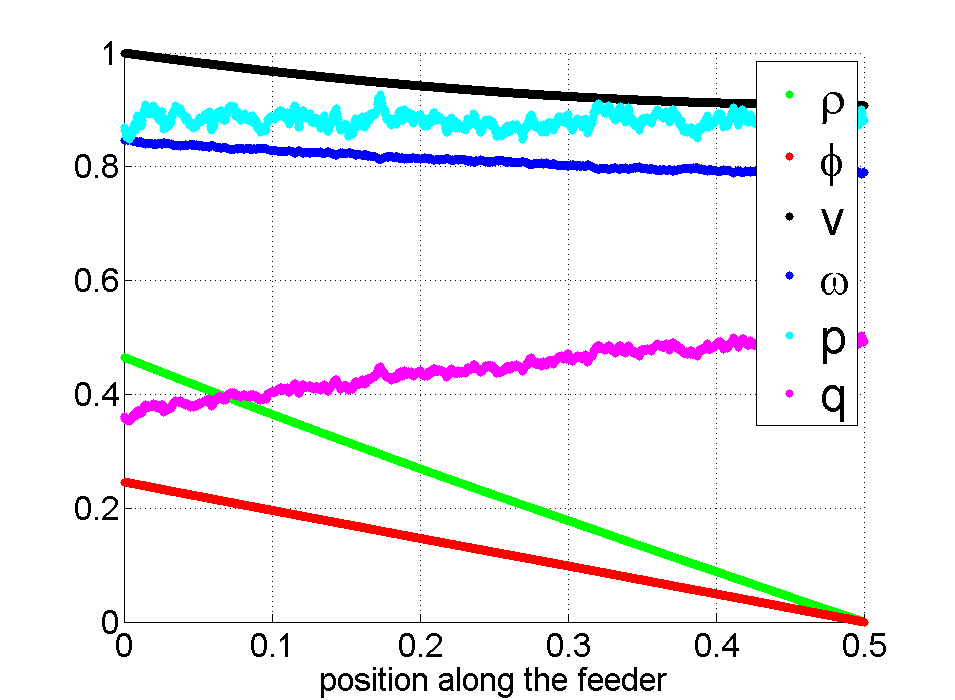}
	\caption{Steady state profile along the feeder for all motors in the ``normal'' state for a sample of disorder drawn from the $\Delta=0.0157$ and $z_d=0.015$ for $v_0=1$. Notice, that while the disorder is clearly seen in active and reactive density consumptions $p$ and $q$, its effect on $\omega$, the flows $\rho$ and $\phi$, and the voltage $v$ is significantly reduced.}
	\label{fig:sample}
\end{figure}

The dynamic simulations described in Section\ref{sec:numerical_approach} are used to find the initial steady profile using two simulation steps.  Dynamic simulation is required because the state of the motors (see Fig.~\ref{fig:hysteresis-single}) is not known {\it a priori}.  Disorder is initially ignored ($\Delta=0$), $v_0$ is set to 1.0, $\omega$ is set to $\omega_0$  and the steady (time-independent) solution of Eqs.~(\ref{p_cont}-\ref{bc}) is found by integrating in time until the solution becomes stationary.  This stationary solution is used as an initial condition for the next simulation.  Using the same $\bar{\tau}_0$, disorder is re-introduced, and Eqs.~(\ref{p_cont}-\ref{bc}) are again integrated until the solution becomes stationary.  Fig.~\ref{fig:sample} shows a typical ``normal'' solution for a sample of disorder drawn from a distribution with $\bar{\tau}_0=0.9$, $\Delta=0.0157$ and $z_d=0.015$.

As clearly seen in Fig.~\ref{fig:sample} the effect of the disorder in $\tau_0$ has little effect on $\omega$ but significantly larger effects on $p$ and $q$. This is obviously the consequence of the structural properties of Eqs.~(\ref{p_cont},\ref{q_cont},\ref{v_eq}). The relatively large effects of disorder on $p$ and $q$ is significantly diminished for $\rho$ and $\phi$ because of the integral relationship between these variables in Eqs.~(\ref{p_cont}, \ref{q_cont}).  The additional integral relationship in Eq.~(\ref{v_eq}) further reduces the effect of disorder on $v$.  It is important to note that the relatively small impact of disorder in Fig.~\ref{fig:sample} is a result of $\omega$ and $v$ being large enough to be far from $v_c^-$, i.e. transition point for high voltage to low-voltage branch in Fig.~\ref{fig:hysteresis-single}.  During the dynamic simulations, this transition region will often reside within the circuit and the relatively small disorder in $\tau_0$ will be magnified by the step change in $p$, $q$, and $\omega$ across this transition.

\subsection{Case A---Stalling Front Dynamics}\label{sec:Stalling}
\begin{figure}
	\includegraphics[width=0.49\linewidth]{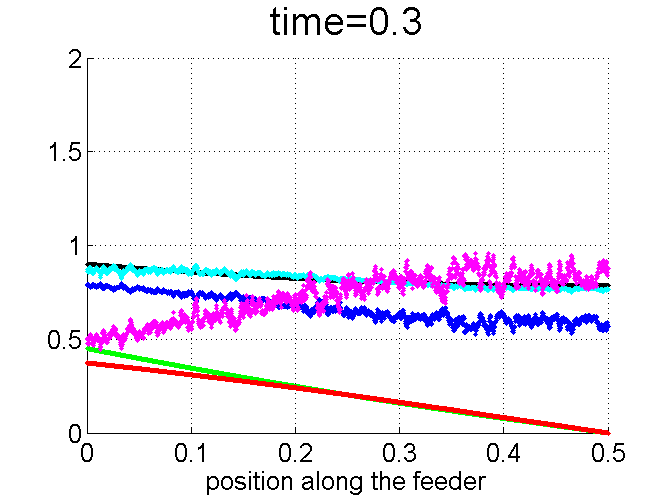}
	\includegraphics[width=0.49\linewidth]{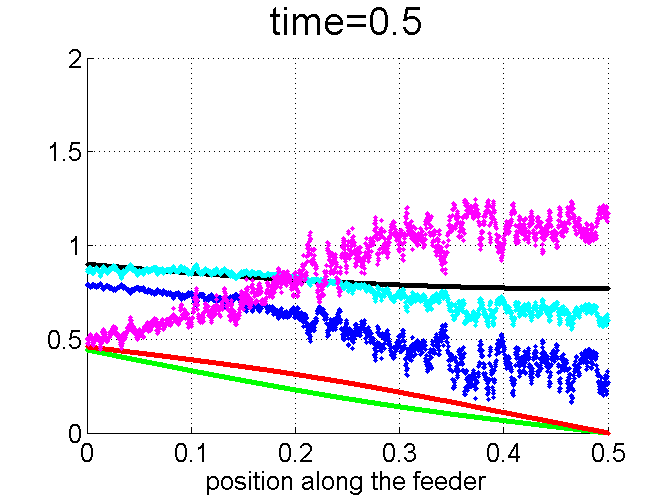}
	\includegraphics[width=0.49\linewidth]{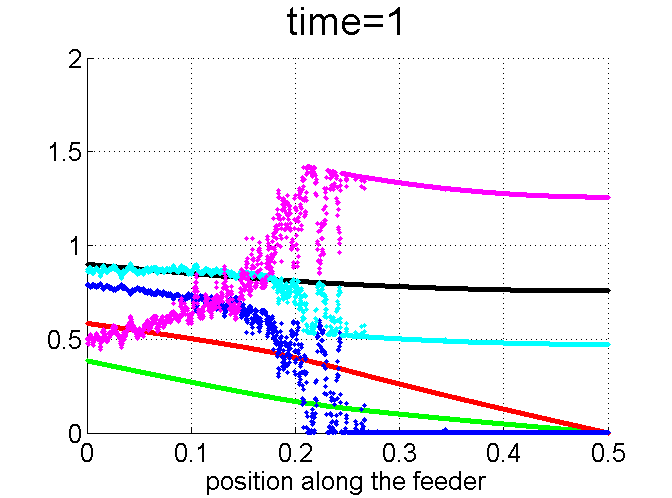}
	\includegraphics[width=0.49\linewidth]{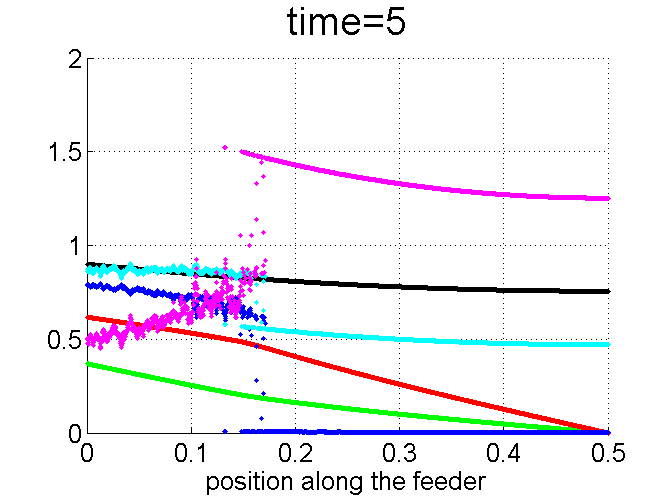}
	\caption{Typical sequence of snapshots of the dynamics for ``Case A---Stalling Front'' for $\Delta=0.0157$ and $z_d=0.005$. The color of the lines is the same as in Fig.~\ref{fig:sample}.  As the stalling front passes from $z=L$ to lower $z$, the disorder in $\tau_0$ interacts with the state transition in Fig.~\ref{fig:hysteresis-single} to amplify the disorder in $\tau_0$.}
\label{fig:snapshotsA}
\end{figure}

Starting from a steady-state condition solved using Case S, $v_0$ is suddenly lowered by a range of $\Delta v$'s to initiate a front of motor stalling.  Figure~\ref{fig:snapshotsA} displays the typical dynamics via a time series of snapshot of $v$, $\rho$, $\phi$, $\omega$, $p$, and $q$.  The reduction of $v_0$ from 1 to 0.9 at $t=0$ lowers $v$ all along the circuit.  The lower voltage reduces the electrical torque on all the motors and the disorder in the mechanical torques causes neighboring motors to decelerate at different rates creating a significant amount of disorder in $\omega$ even before any of the motors crosses the transition from a normal to a stalled state (see Fig.~\ref{fig:snapshotsA} at $t$=0.3).  At $t=0.5$, motors near the end of the line begin to stall, however, they all begin this transition at slightly different times because of the initial variability in their deceleration rates.  The rapid deceleration during this transition significantly amplifies the disorder $\omega$ caused by the disorder in $\tau_0$  The disorder also appear in $q$ because of the large difference in $q$ on either side of the transition.  At $t=1$, the the motors near the end of the circuit have completed their stalling transition and are all near $\omega=0$ which suppresses the impact of the disorder in $\tau_0$.  The disorder in $\tau_0$ is only amplified near the stalling front where significant dynamics are still occurring.  At $t=5$, the dynamics have essentially ceased. The disorder in $\tau_0$ causes each motor to have slightly different $v_c^-$ (see Fig.~\ref{fig:hysteresis-single}, and we believe that this variation in the transition threshold causes the majority of the residual disorder in $\omega$ and $q$.

\begin{figure}
		\includegraphics[width=1\linewidth]{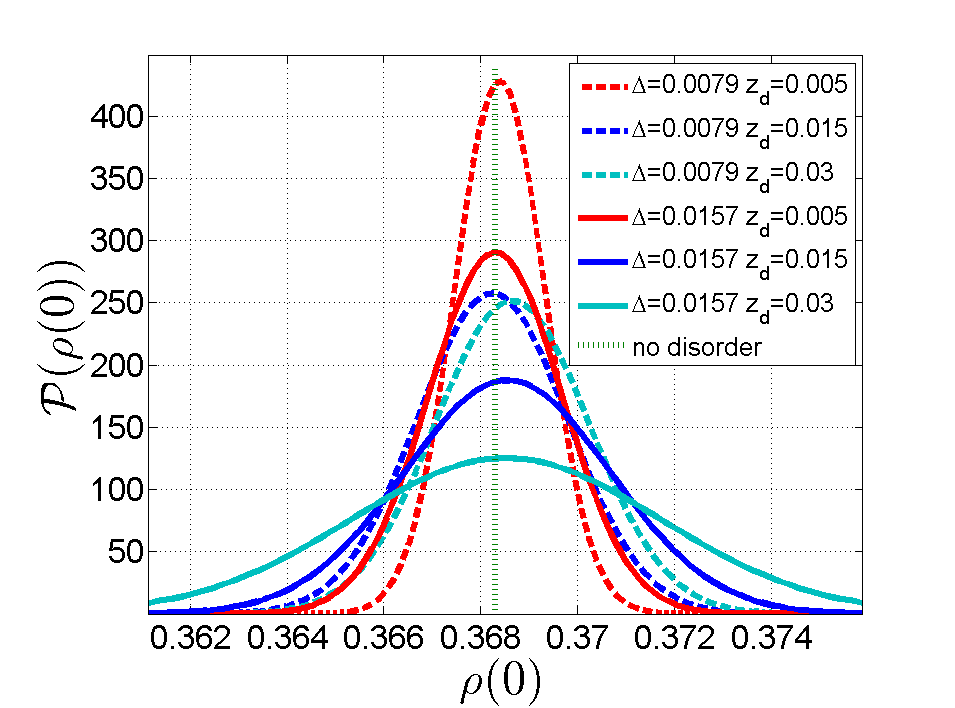}
		\includegraphics[width=1\linewidth]{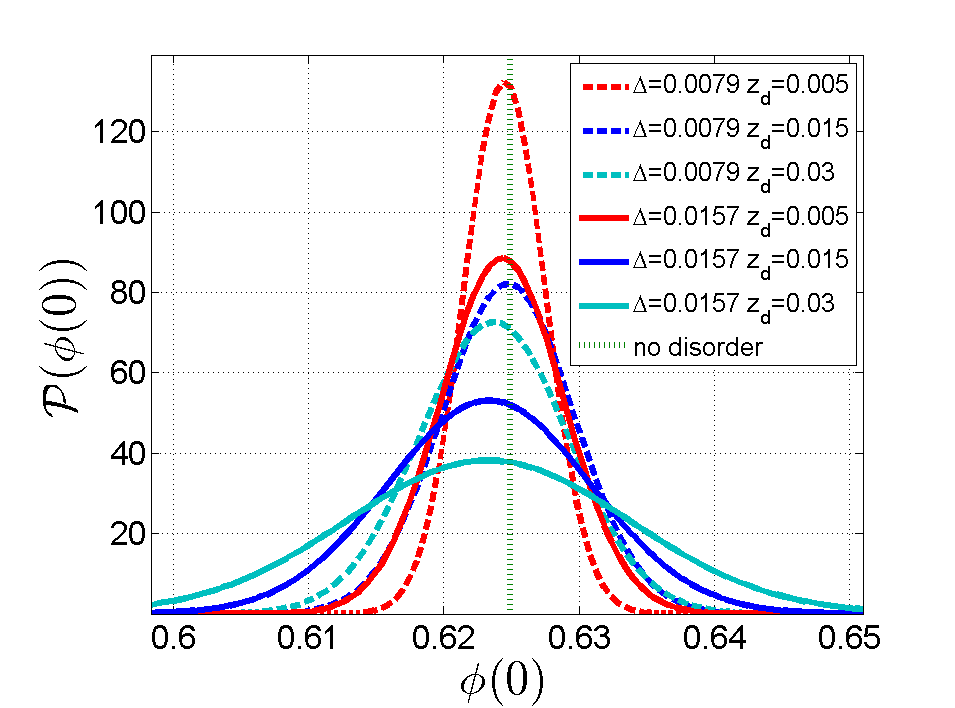}
		\includegraphics[width=\linewidth]{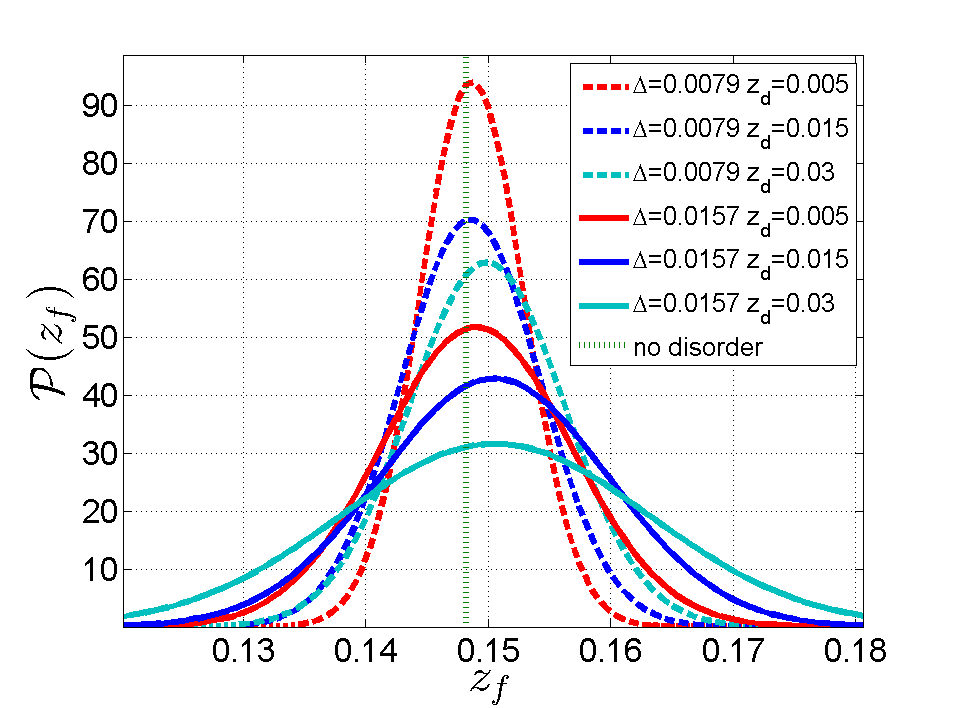}
	\caption {Gaussian fits to the probability density distribution functions over $\rho(0)$, $\phi(0)$ and $z_f$  after reaching steady state for Case A---Stalling front dynamics. Statistics are collected over 100 samples each disorder ensemble parameterized by $\Delta$ and $z_d$.  The vertical dashed line indicates the position of the front for no disorder, i.e. $\Delta$=0 and $z_d$=0.   } \label{fig:caseA}
\end{figure}

The same reduction in $v_0$ is applied to one hundred samples of several different ensembles of disorder, i.e. different values of $\Delta$ and $z_d$.  For each ensemble, the steady state values of $\rho(0)$, $\phi(0)$, and $z_f$ are collected and binned into histograms.  Gaussians are fit to these histograms, and the results are plotted in Fig.~\ref{fig:caseA}.  We first consider the behavior of $z_f$, i.e. the probability distribution $\mathcal{P}(z_f)$. For each sample the value of $z_f$ is found via the average between the most left point in the "stalled" state and the most right one in the "normal" throughout the feeder. For a given correlation length $z_d$ of the $\tau_0$ disorder, the width of $\mathcal{P}(z_f)$ grows as $\Delta$ grows.  (See the difference in the solid and dashed traces of the same color in Fig.~\ref{fig:caseA}).  In fact, for each $z_d$, the width of $\mathcal{P}(z_f)$ approximately doubles for a doubling in $\Delta$.

This effect appears to be local. Specifically, if $\Delta$ was zero, the front would stop in the same place $z_f^0$ for each sample.  With variability in $\tau_0$, i.e. $\Delta \neq 0$, the front might stall slightly earlier ($z_f>z_f^0$) if it encounters a small cluster of low $\tau_0$ motors at slightly larger $z_f$.  Or, it may stall slightly later ($z_f<z_f^0$) if there is a small cluster of high $\tau_0$ motors near $z_f^0$ with cluster of low $\tau_0$ motors at smaller $z$.  The finite correlation length $z_d$ ensure that such clusters will exist.  These correlations could create nonlocal effects of disorder, however, we do not believe this is the case.  In Figs.~\ref{fig:snapshotsA} and \ref{fig:caseA}, the $z_f$ are distant from the end of circuit compared to $z_d$.  The relatively small $z_d$ and the smoothing discussed in Section~\ref{sec:Case S} drastically reduce any residual effect of the disorder in $\tau_0$ on the voltage profile with the result that the average position of $z_f$ is nearly the same in all cases.  However, at the larger $z_d$, there does appear to be a slight shift toward larger $z_f$ although this result is not definitive.

The general trend is the same for the other two variables $\rho(0)$ and $\phi(0)$.  However, this is expected because the final location of stalled front has a major influence over the motor loads which in turn create both $\rho(0)$ and $\phi(0)$.

\begin{figure}
	\centering
	\includegraphics[width=1\linewidth]{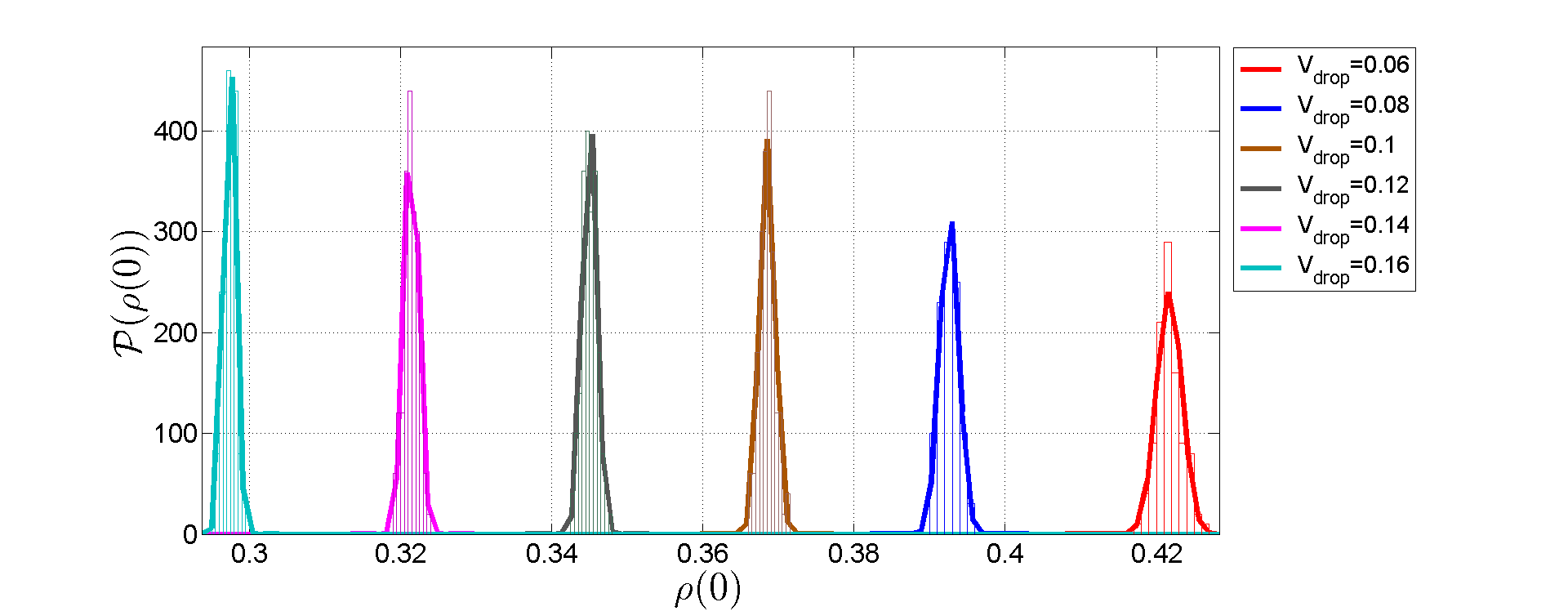}
	\includegraphics[width=1\linewidth]{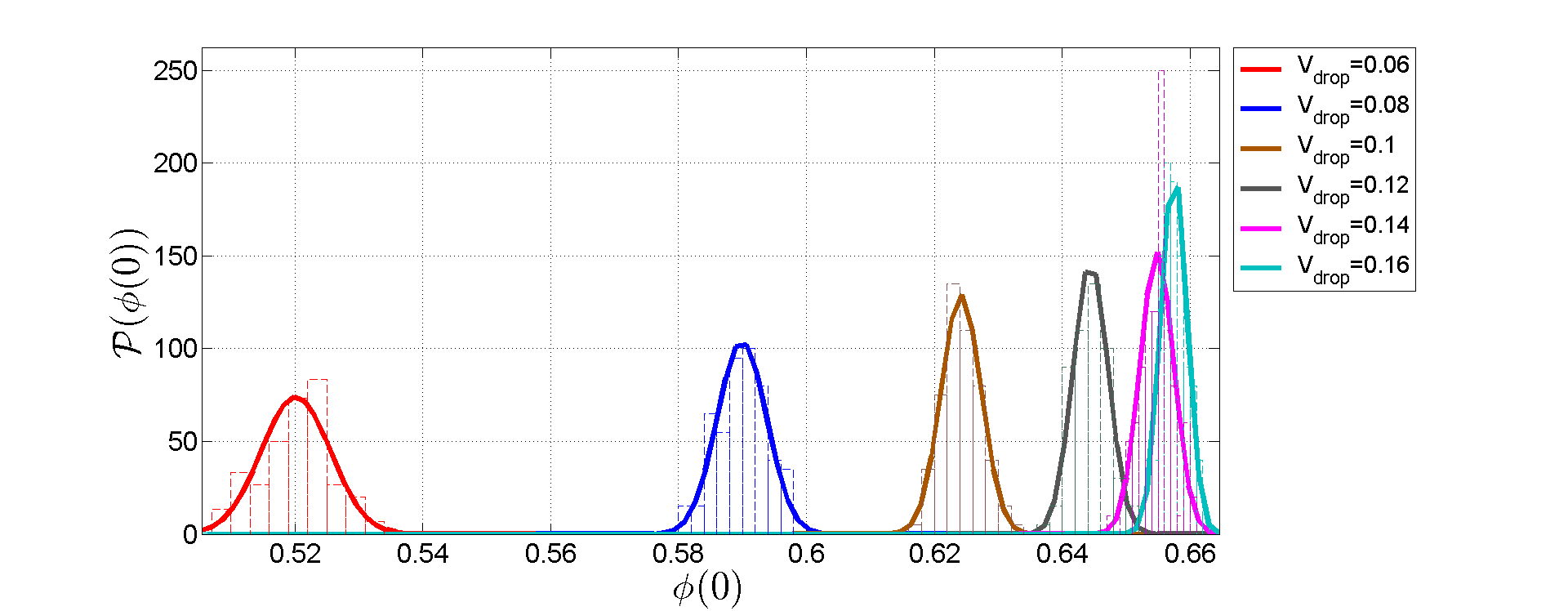}
	\includegraphics[width=1\linewidth]{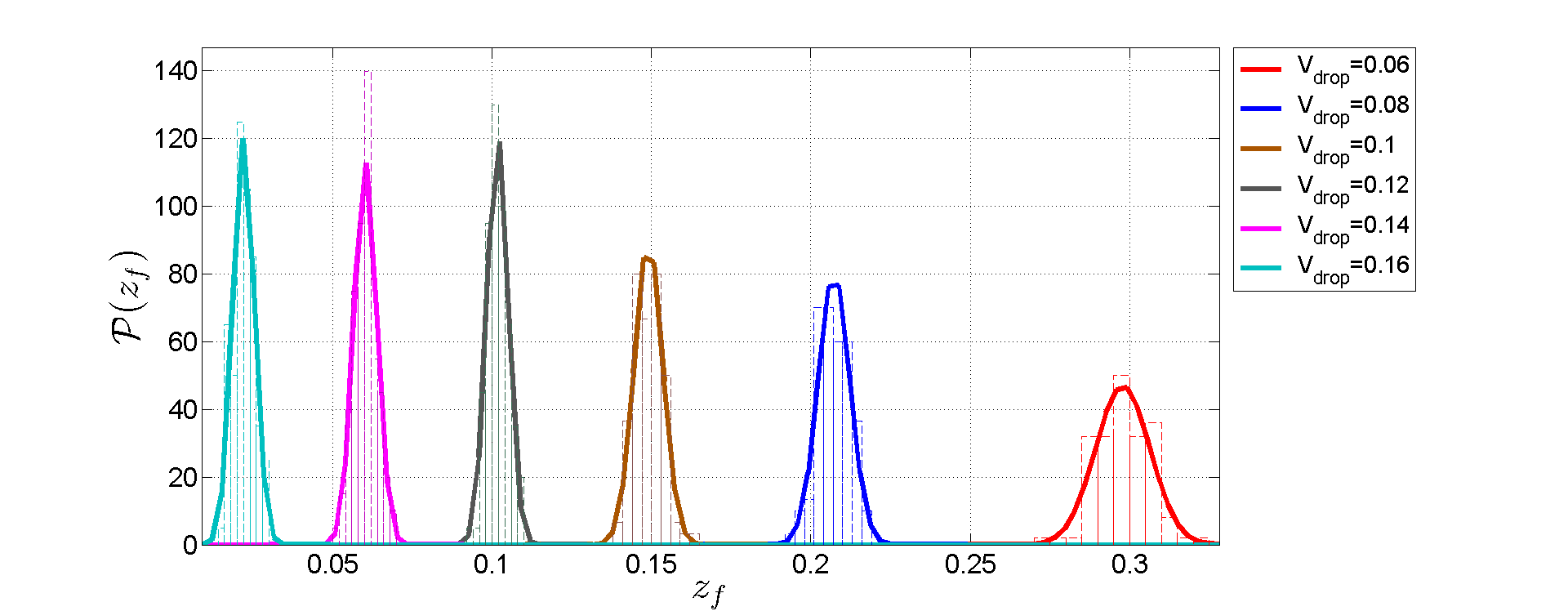}
	\caption{Probability distribution functions over $\rho(0)$, $\phi(0)$ and $z_f$ observed for the $\Delta=0.0079$ and $z_d=0.005$ ensemble after reaching steady state following Case A---Stalling front dynamics for different values of $v_0$. }
\label{fig:volt_drop_change}
\end{figure}

Additional numerical experiments shed more light on the effects of disorder in $\tau_0$ on the properties of the stalled front.  In Fig.~\ref{fig:volt_drop_change}, simulations are performed for a fixed disorder ensemble ($\Delta$=0.0079 and $z_d=0.005$), but the reduction is $v_0$ is varied.  As expected, for smaller $v_0$ the front stalls closer to the end of the circuit.  However, the width of $\mathcal{P}(z_f)$ is larger than when the front stalls at smaller $z$ (i.e. for larger reductions in $v_0$).  The cause of this larger width is two-fold.  First, the slope of $v(z)$ is smaller for $z\sim L$ making the position of the front much more susceptible disorder.  Second, being closer to the end of the circuit provides less smoothing of $v$, and the correlation length $z_d$ is more effective at causing variations in $v$ near to the nominal location of stalling.  The effects on  $\mathcal{P}(\rho(0))$ and $\mathcal{P}(\phi(0))$ in Fig.~\ref{fig:volt_drop_change} can be inferred from $\mathcal{P}(z_f)$ and Eqs.~\ref{p_cont}-\ref{q_cont}.

\subsection{Case B---Restoration Front Dynamics}

\begin{figure}

	\includegraphics[width=0.49\linewidth]{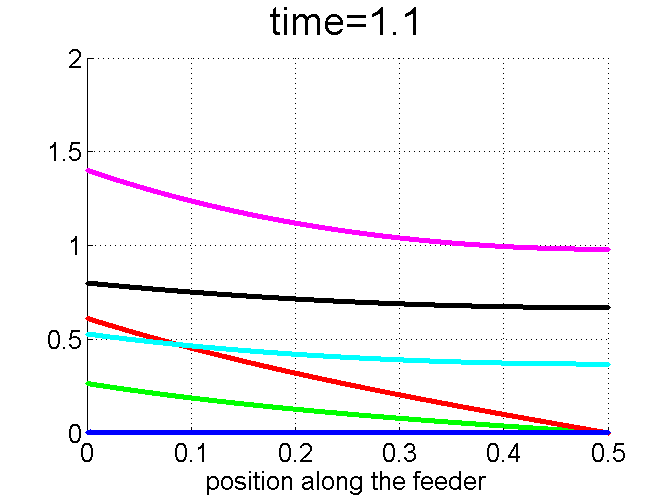}
	\includegraphics[width=0.49\linewidth]{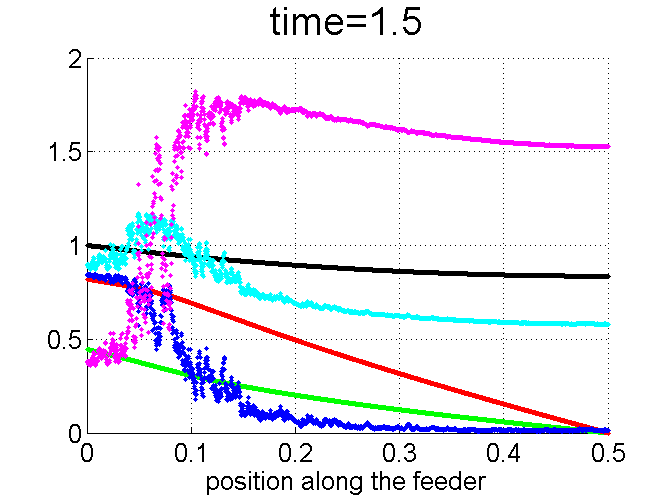}
	\includegraphics[width=0.49\linewidth]{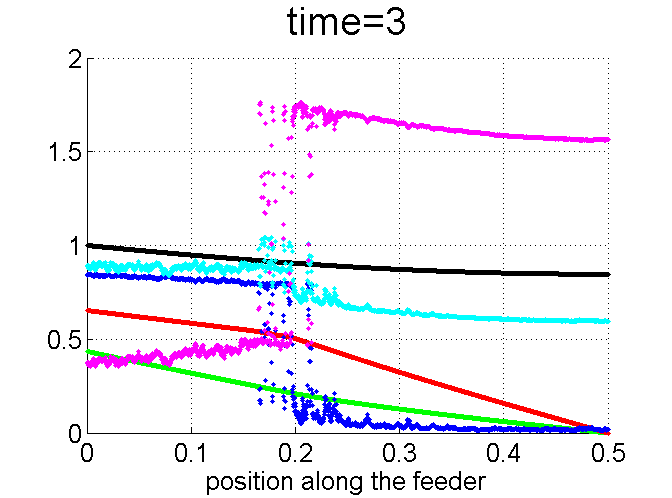}
	\includegraphics[width=0.49\linewidth]{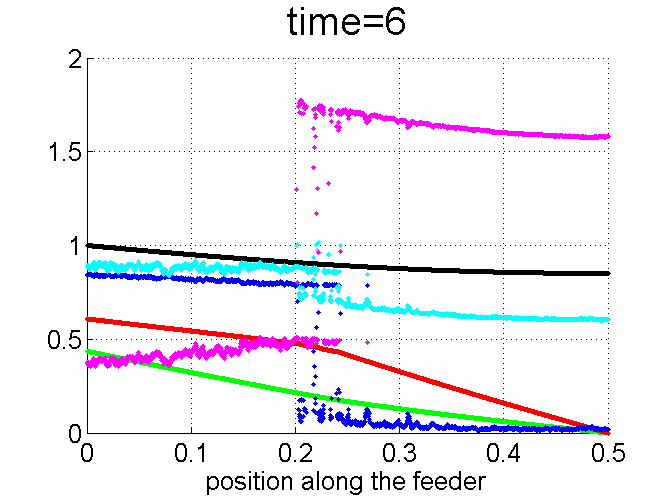}
	\caption{Typical sequence of snapshots of the dynamics for ``Case B---Restoration Front'' for $\Delta=0.0157$ and $z_d=0.005$. The color of the lines is the same as in Fig.~\ref{fig:sample}.  As the restoration front passes from $z=0$ into the circuit, the disorder in $\tau_0$ interacts with the state transition in Fig.~\ref{fig:hysteresis-single} to amplify the disorder in $\tau_0$ .}
\label{fig:snapshotsB}
\end{figure}

Fig.~\ref{fig:snapshotsB} show the snapshots of the dynamics for a restoration front.  At $t$=1.1, $v_0$=0.8 and has been held at this low value long enough so that all of the motors on circuit have stalled.  The disorder in $\tau_0$ has little effect because all of the motors are stalled and are far from $v_c^+$, i.e. the transition to the normal state. Immediately following $t$=1.1, $v_0$ is raised back to 1.0 launching a recovery front into the circuit from $z$=0. The effect of the disorder is very similar to the stalling front.  Specifically, at $t$=1.5, the motors beyond the front are just beginning to accelerate creating moderate disorder in $\omega$ because of the different rates of acceleration.  Within the front, this disorder is amplified during the rapid dynamical transition from the stalled to the normal state.  For locations behind the front, i.e. small $z$, the disorder has limited effect on $\omega$ because the accelerations have mostly ceased.  This behavior persists though the simulation up to $t$=6 when the front is nearly stationary.  At these long times, the effect of the disorder is again local and static, i.e. disorder in $\tau_0$ drives disorder in $v_c^+$ for each motor which manifests as randomness in the motor state near the stall restoration front.

\begin{figure}
		\includegraphics[width=\linewidth]{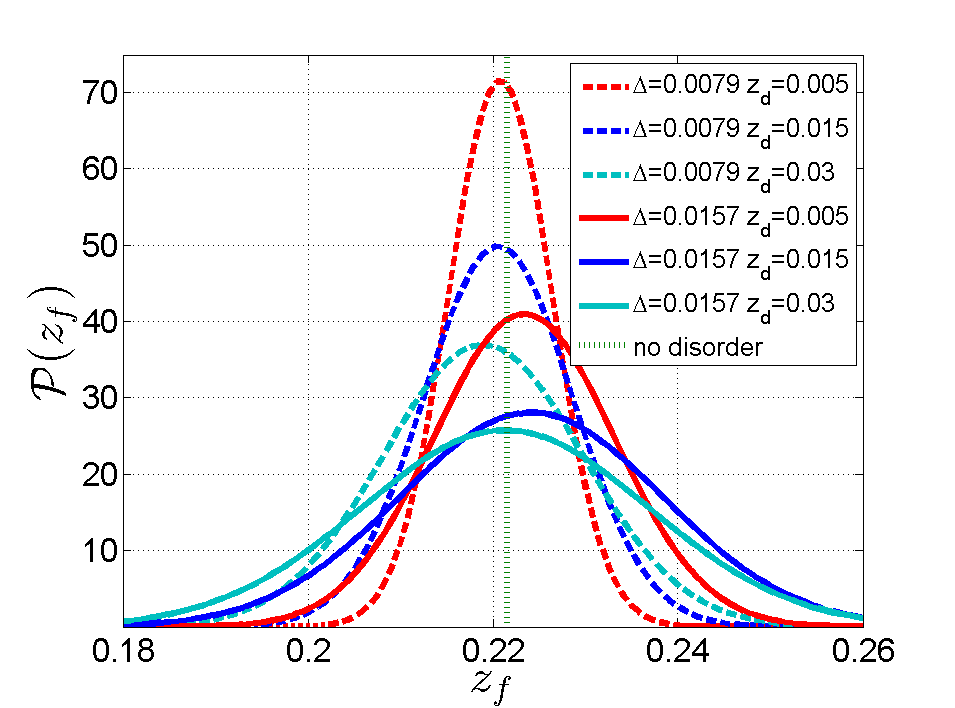}
	\caption {Gaussian fits to the probability density distribution function over $z_f$  after reaching steady state for Case B---Restoration front dynamics. Statistics are collected over 100 samples each disorder ensemble parameterized by $\Delta$ and $z_d$. The vertical dashed line indicates the position of the front for no disorder, i.e. $\Delta$=0 and $z_d$=0.  } \label{fig:caseB}
\end{figure}

The probability distribution $\mathcal{P}(z_f)$ for the steady-state $z_f$ is shown in Fig.~\ref{fig:caseB}.  The effect of disorder on $\mathcal{P}(z_f)$ is very similar to Case A--Stalling front dynamics.  At fixed $z_d$, the width of $\mathcal{P}(z_f)$ approximately doubles when the amplitude of the disorder $\Delta$ doubles. Following a similar argument as in Section~\ref{sec:Stalling}, we conclude that the effect of the disorder is primarily local and somewhat contingent on $z_d$, i.e. pockets of motors with high and low $\tau_0$ that catch the restoration front early or allow it to propagate a bit further before becoming stationary.  Fig.~\ref{fig:caseB} does not show $\mathcal{P}(\rho(0))$ or $\mathcal{P}(\phi(0))$, but their relationship to $\mathcal{P}(z_f)$ is very similar to the relationship in Fig.~\ref{fig:caseA}.

\subsection{Case C---Fault Clearing}\label{sec:fault_clearing}

The dynamics of fault clearing is more complex than just a stalling front or a restoration front.  In fault clearing, the circuit starts out in a steady state with $v_0$=1 and all of the motors in the normal state (see Fig.~\ref{fig:hysteresis-single}).  At $t$=0, the $v_0$ is reduced to 0.9 and held low for time $\tau$.  After $\tau$, $v_0$ is restored to 1.0 and the dynamics are simulated until the motors reach a steady state. A bisection search in $\tau$ is used to find $\tau_{cl}$, i.e. the maximum clearing time where the motors at the end of the circuit will just recover. The circuit is considered ``restored'' even if there are just a few locations ($\sim$ 1-3) with stalled motors.  The search is carried out for 100 realizations of four disorder ensembles.

\begin{figure}
	\includegraphics[width=0.95\linewidth]{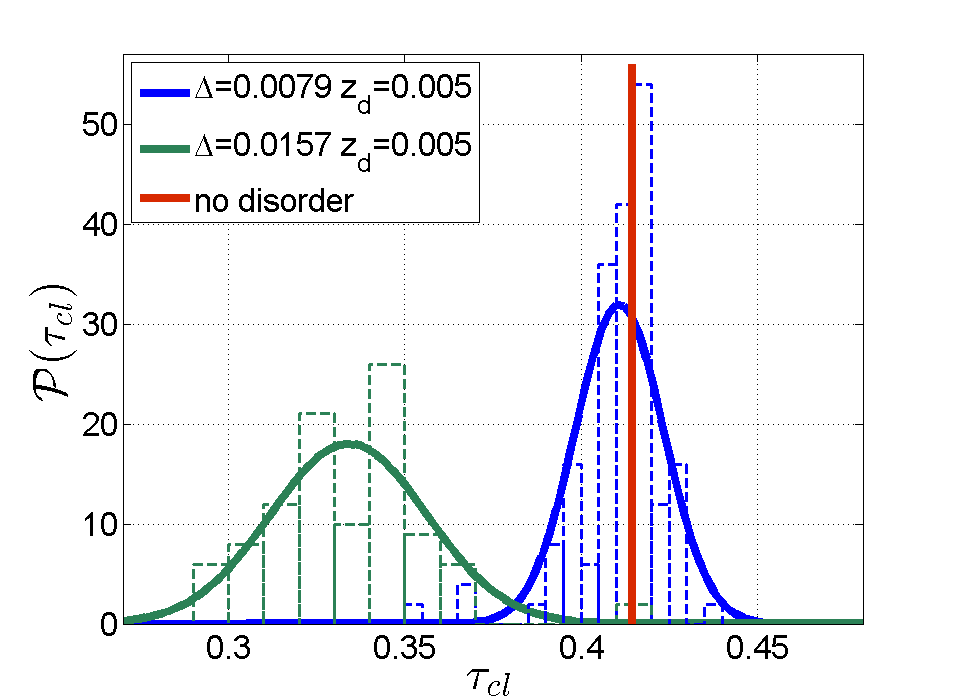}
	\includegraphics[width=0.95\linewidth]{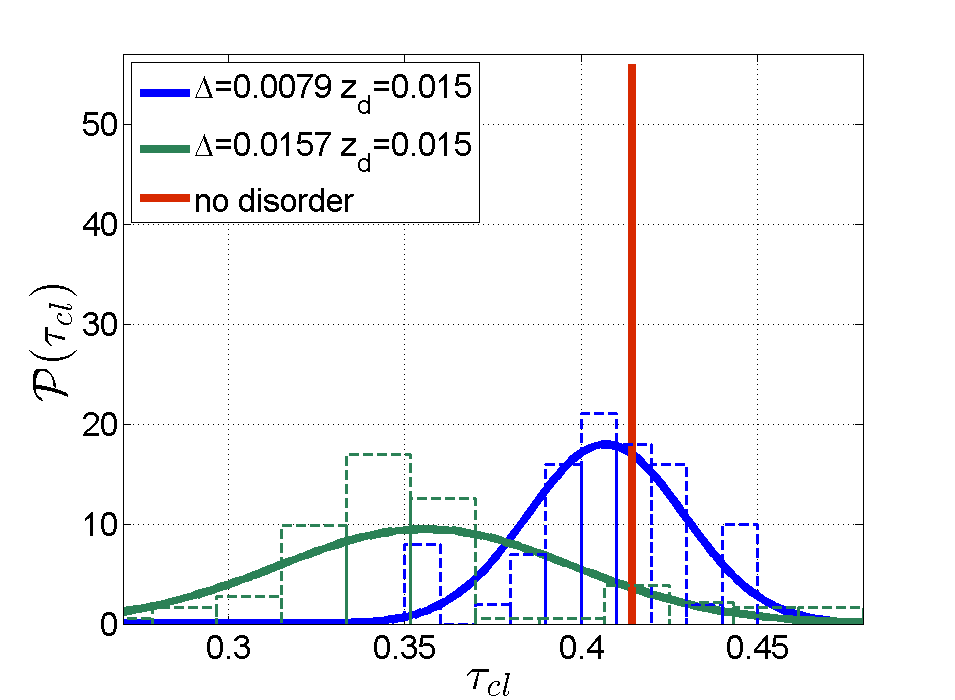}
	\caption{ Histograms and respective Gaussian fits for the probability density distribution function of the maximum clearing time, $\tau_{cl}$, measured under experiments C for four different ensembles and for the case of no disorder (vertical red lines).  }
\label{fig:tau_clearing}
\end{figure}

The distributions of $\tau_{cl}$, i.e. $\mathcal{P}(\tau_{cl})$ are plotted in Fig.~\ref{fig:tau_clearing}.  In general, an increase in the amplitude of the disorder $\Delta$ results in a shift of $\mathcal{P}(\tau_{cl})$ to shorter $\tau_{cl}$ while an increase in the correlation length $z_d$ of the disorder results in significant broadening of $\mathcal{P}(\tau_{cl})$.  The complexity of the fault-clearing dynamics become evident by comparing the maximum clearing times in Fig.~\ref{fig:tau_clearing} with the snapshots of a stalling front in Fig.~\ref{fig:snapshotsA}.  The $\tau_{cl}$ are on the order of 0.3 to 0.5 in Fig.~\ref{fig:tau_clearing}.  Inspection of Fig.~\ref{fig:snapshotsA} at these times shows that the stalling front has not yet reached a steady state.  In fact, none of the motors has even reached $\omega/\omega_0\sim$ 0. Before attempting to understand the effects of disorder on $\tau_{cl}$, we first give a qualitative description of the fault-clearing dynamics.

A qualitative understanding of boundary between a ``stalled'' circuit, i.e. the occurrence of a FIVDR event, and a ``recovered'' circuit is gained by inspecting the snapshots of the dynamics in Fig.~\ref{fig:snapshotsA}.  This simulation corresponds to disorder parameterized by $\Delta$=0.0157 and $z_d$=0.005, which is the same as the green curve in the upper plot of Fig.~\ref{fig:tau_clearing}.  Consider the $t$=0.5 snapshot in Fig.~\ref{fig:snapshotsA}.  The motors near the end of the circuit have $v\sim$ 0.75  and $\omega/\omega_0\sim$ 0.3.  If $v_0$ was restored to 1.0 at this time, the end of the line would at best have $v\sim$ 0.85. In reality, it would be lower because the reactive power consumption of the motors would increase at the higher voltage.  However, mapping the state $v\sim$ 0.85 and $\omega/\omega_0 \sim$0.3 onto the torque plot of Fig.~\ref{fig:torques}, we find that the electrical torque falls below the mechanical torque.  Therefore, even after the fault is cleared and $v_0$ is restored to 1.0, the motors at the end of the circuit will continue to decelerate.  As they slow, their reactive power increases somewhat (see Eq.~\ref{q_cont}) which has a tendency to suppress the voltage further.  The result is that, at $t$=0.5, the motors near the end of the circuit will continue to decelerate to near $\omega/\omega_0\sim$ 0 even after the fault is cleared---a conclusion consistent with the clearing time plots in Fig.~\ref{fig:tau_clearing}.

At $t$=0.3 in Fig.~\ref{fig:snapshotsA}, the situation is very different.  The motors near the end of the circuit have only decelerated to $\omega/\omega_0 \sim$ 0.6 and the local voltage is $v\sim$ 0.8.  If the fault was cleared at $t$=0.3, the voltage at the end of the circuit was jump up to about 0.9.  Mapping  the post-fault clearing state $\omega/\omega_0 \sim$ 0.6 and $v\sim$0.9 onto the torque curves in Fig.~\ref{fig:torques}, we find that the electrical torque is safely above the mechanical torque, and even the motors at the end of the circuit is begin to accelerate after the fault is cleared.  At higher $\omega/\omega_0$, their reactive power consumption decreases which reinforces the increase in voltage and the overall recovery.

From this qualitative description, we expect that the effect of disorder is primarily felt in the initial deceleration of the motor while the fault is applied rather than during post-fault recovery period.  The spatially correlated disorder will result in clumps of motors with higher than average $t_0$.  These motors will decelerate faster than an average motor pushing them to lower values of electrical torque along a constant $v$ curve in Fig.~\ref{fig:torques}.  The tendency is for these motors to experience a decelerating net torque after fault clearing.  The implication is that average maximum clearing times $\tau_{cl}$ become shorter and more broadly distributed.

\subsection{Width of the blurry region}

\begin{figure}
	\centering
	\begin{minipage}{0.49\linewidth}
		\centering
		\includegraphics[width=1\linewidth]{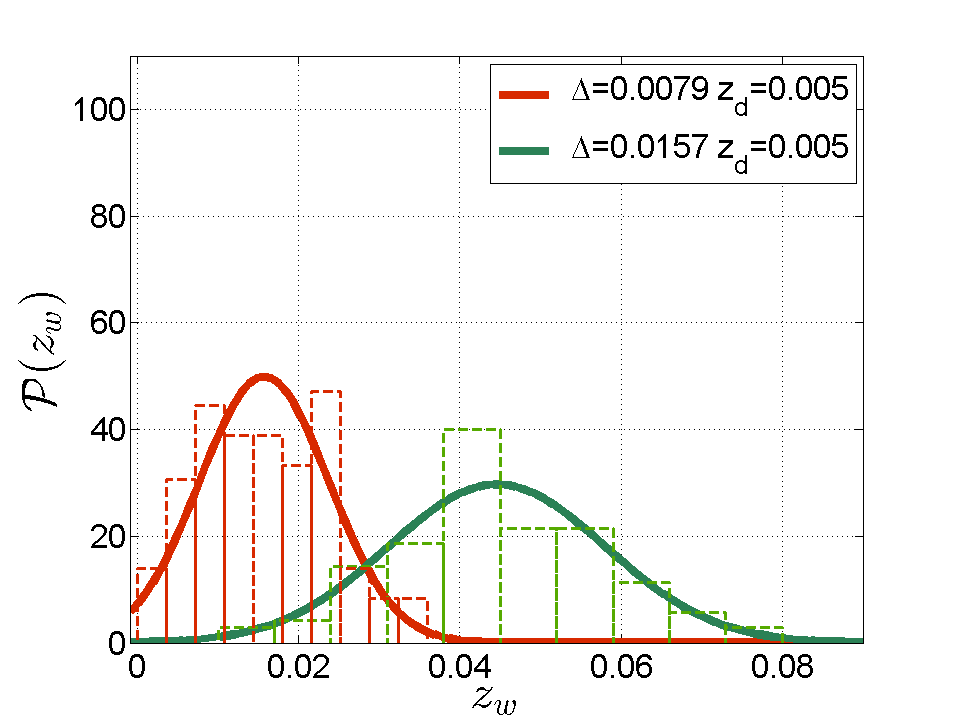}
		\includegraphics[width=1\linewidth]{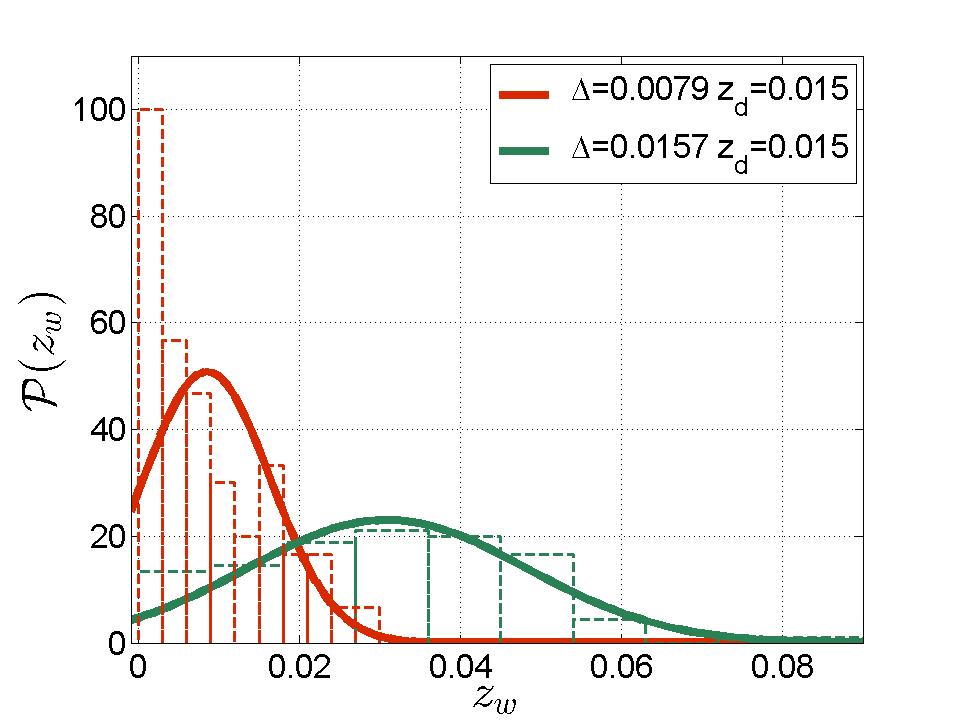}
		Case A
	\end{minipage}
	\begin{minipage}{0.49\linewidth}
		\centering
		\includegraphics[width=1\linewidth]{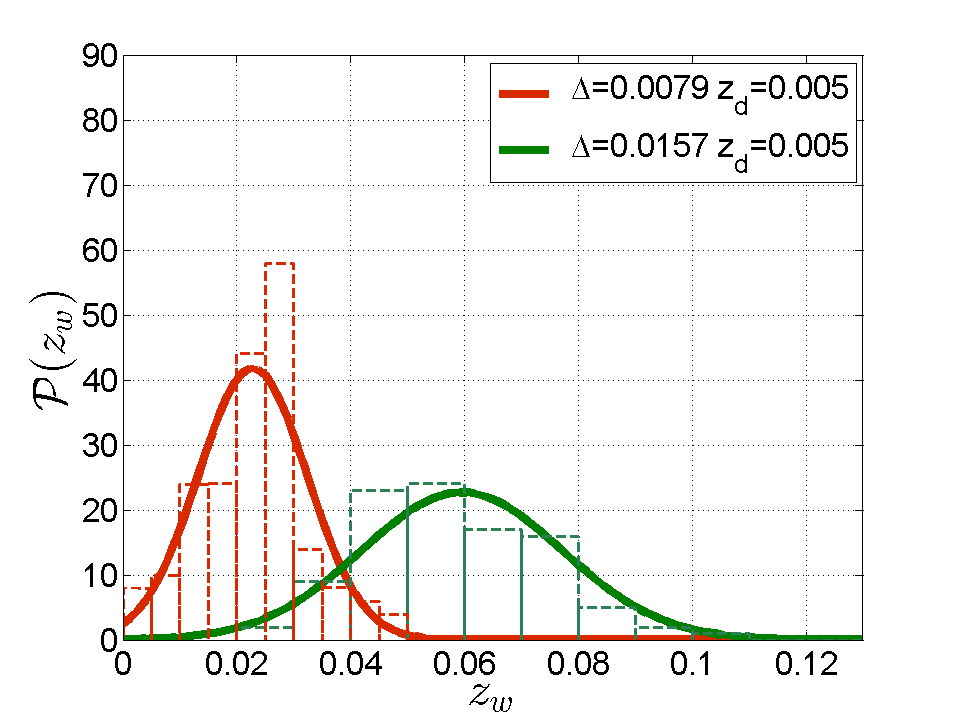}
		\includegraphics[width=1\linewidth]{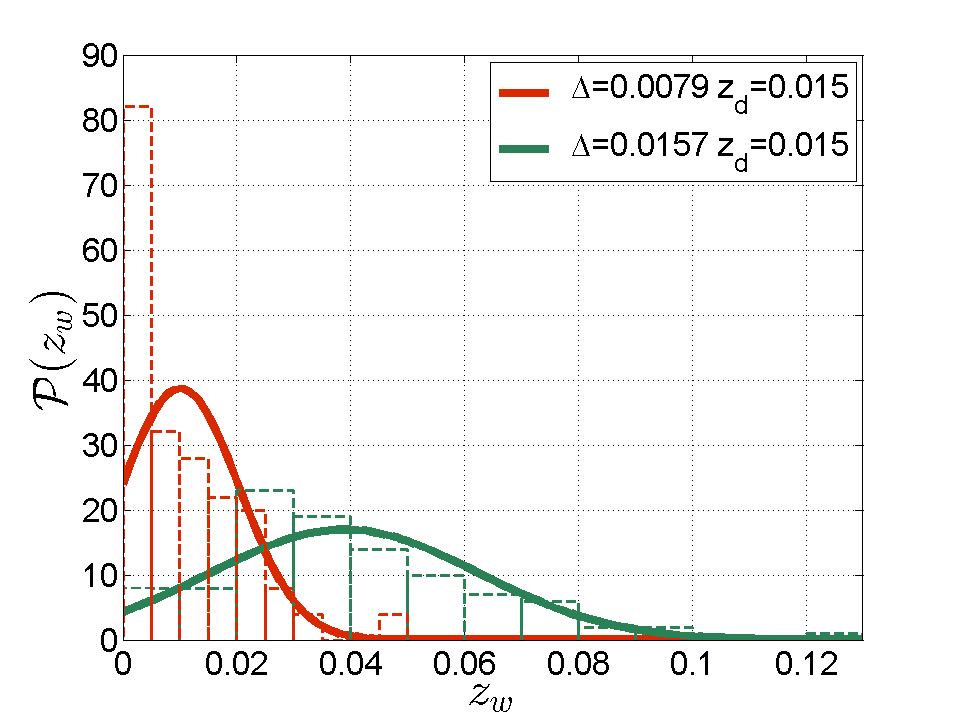}
		Case B
	\end{minipage}
	\caption{Histograms and respective probability density distribution functions for the width of the burry regions, $z_w$, shown for the final state of the respective ensembles (see the legend) in experiments A and B.}
\label{fig:z_w}
\end{figure}

An interesting consequence of the disorder is seen in Fig.~\ref{fig:z_w} where the probability density distribution function of the blurry region width, $z_w$ is shown for different values of the disorder amplitude $\Delta$ and correlation length, $z_d$. We observe, in particular, that $z_w$ increases with increase in $\Delta$ (when $z_d$ is fixed) and decrease in $z_d$ (when $\Delta$ is fixed). We postpone detailed discussions and explanations of this phenomenon for future publications.

\section{Conclusions and Path Forward}\label{sec:Conclusions}

Loading on electrical distribution circuits is far from uniform and is often clumped into load pockets distributed along the circuit.  To better represent the effect of these conditions on the distribution grid dynamics of induction motor loads, we have introduced an electrical load model that includes spatially-correlated load disorder.  To investigate the effects of this new loading, we have performed numerical simulation of the dynamics of a radial distribution circuit using this model of load disorder and explored the effect of disorder on the critical clearing time to avoid a Fault-Induced Delayed Voltage Recovery (FIDVR) event. Although the effects of disorder do bring new and  important qualitative behaviors, by and large, the main qualitative picture of the front propagation phenomena observed in the spatially homogeneous case\cite{13DCB} remains the same but with some differences in the more complex dynamics of fault clearing.  Specifically,
\begin{itemize}
\item For both stalling and restoration dynamics, the fronts propagate in time, slow down and eventually stop in a partially stalled state (for the right combinations of circuit length and voltage perturbation)
\item For relatively small disorder, there is a threshold, i.e. a reasonably well defined maximum clearing time $\tau_{cl}$, that separates the final circuit states into fully restored ($\tau < \tau_{cl}$) and only partially restored ($\tau > \tau_{cl}$).
\item However, as the disorder becomes larger in amplitude with longer correlation lengths, the distribution of maximum clearing times becomes quite broad.
\end{itemize}

The broad distribution of maximum clearing times is likely related to new qualitative effects that emerge from the presence of disorder. Specifically, Figs.~\ref{fig:snapshotsA} and \ref{fig:snapshotsB} both show that a group of motors with a distribution of mechanical torques undergoing acceleration or deceleration acquire a wide distribution of motor rotational frequencies.  This effect is particularly evident in Fig.~\ref{fig:snapshotsA} at $t$=0.5.  Motors with higher mechanical torque undergoing deceleration during a fault reach lower rotational frequencies and are in a more precarious situation.   After fault clearing, they may not recover to normal rotational rate near grid frequency.  Instead they may experience a net decelerating torque and stall.  The effect of this local stalling on surrounding motors is still an unresolved questions.

There are many ways that this work could be extended and improved, including:
\begin{itemize}
    \item Improving the load models by including spatially-distributed constant impedance, constant current, or constant power loads and investigating the effects of these combined loads on the induction motor dynamics.
    \item The exploration of analytical approximations to the maximum fault clearing time based on the qualitative description of post-fault recovery in Section~\ref{sec:fault_clearing}.
    \item Extension of the model to distribution circuits with multiple branches and/or multiple circuits emanating from a single substation.
    \item The development of new controls to arrest a FIDVR event before it becomes established, possibly using distributed control of reactive power generation by customer-owned inverters \cite{PV_inverter_Q_2011}.
\end{itemize}


\begin{acknowledgments}
The work at LANL was carried out under the auspices of the National Nuclear Security Administration of the U.S. Department of Energy at Los Alamos National Laboratory under Contract No. DE-AC52-06NA25396. MC and SB also acknowledge partial support of the Advanced Grid Modeling Program in the US Department of Energy Office of Electricity and of the
NSF/ECCS collaborative research project on Power Grid Spectroscopy through NMC.
\end{acknowledgments}

\bibliographystyle{unsrt}
\bibliography{Bib/FIDVR,Bib/SmartGrid,Bib/voltage}

\end{document}